# Black Magic Investigation Made Simple: Monte Carlo Simulations and Historical Back Testing of Momentum Cross-Over Strategies Using FRACTI Patterns

Jorge M. Faleiro Jr [#1], Edward P. K. Tsang [#2]

[#] Centre of Computational Finance and Economic Agents (CCFEA), University of Essex
Wivenhoe Park, Colchester, CO4 3SQ, UK
[1] jfalei@essex.ac.uk
[2] edward@essex.ac.uk

*Abstract—* **To promote economic stability, finance should be studied as a hard science, where scientific methods apply. When a trading strategy is proposed, the underlying model should be transparent and defined robustly to allow other researchers to understand and examine it thoroughly. Any report on experimental results must allow other researchers to trace back to the original data and models that produced them. Like any hard sciences, results must be repeatable to allow researchers to collaborate, and build upon each other's results. Large-scale collaboration, when applying the steps of scientific investigation, is an efficient way to leverage "crowd science" to accelerate research in finance.**

**In this paper, we demonstrate how a real world problem in economics, an old problem still subject to a lot of debate, can be solved by the application of a crowd-powered, collaborative scientific computational framework, fully supporting the process of investigation dictated by the modern scientific method.**

**This paper provides a real end-to-end example of investigation to illustrate the use of the framework. We intentionally selected an example that is self-contained, complete, simple, accessible, and of constant debate in both academia and the industry: the performance of a trading strategy used commonly in technical analysis. Claims of efficiency in technical analysis, referred derisively by some sources as "Black Magic", are of widespread use in mainstream media and usually met with a lot of controversy.**

**In this paper we show that different researchers assess this strategy differently, and the subsequent debate is due more to the lack of method than purpose. Most results reported are not repeatable by other researchers. This is not satisfactory if we intend to approach finance as a hard science.**

**To counterweight the** *status quo***, we demonstrate what one could do by using collaborative and investigative features of contributions and leveraging the power of crowds.**

## I. Crowds and Computational Investigation

Finance is not always studied with rigor. Even the simplest and most popular trading strategies are disputed over their performance. This is not just because the different researchers use different data to test the strategies, but also because contexts are not always clearly defined, parameters are not always agreed, and results are not always interpreted in the same way. The stability of financial markets affects the society; many societies are still suffering from the effect of the financial crisis in 2007-08. To promote economic stability, finance should be studied as a hard science, where scientific methods apply.

In this paper we present a real end-to-end example of an applied investigation of a financial trading strategy using a specialized *conceptual framework*. We show how this framework can be used to leverage the power of the crowds, modern computing resources [1] and the scientific method [2] to solve practical problems in economics and financial markets through large-scale collaboration [1].

This paper is part of a broad research in which we advocate a "crowd powered" [3], computer-based model of investigation for complex subjects like economics. The specific intent of this present paper is to serve as a continuation for the blueprint of the conceptual framework [1] already published, offering a concrete example of an end-to-end investigation of a financial case using constructs of this conceptual framework. We consider this paper an important formalization that will help consolidate the vision of the conceptual framework.

We have been calling this conceptual framework **FRACTI**: FRAmework for Collaboration and Transparent Investigation in Financial Markets [1]. FRACTI is not a software implementation, or a programming language. Instead, it defines an abstraction for a computational representation [4] for the field of economics. The examples outlined in this paper denote one specific implementation, or dialect, of FRACTI (an open source implementation called QuantLET [5]). This dialect evolved to serve as an illustration of concepts in the framework and is not intended to be a full-fledged implementation of all concepts at this point.

Throughout this work the dialect has been providing important insights to this research, and vice-versa, the research fed back to the dialect several ideas that were materialized as concrete extensions. This dialect relies on many underlying computational resources [6] [7] [8] [9] and we should expect more to come, as the chain of direct and indirect dependencies is fluid and is always changing. It would be impossible to accurately describe all of them.

We will be using FRACTI models to investigate whether one common trading strategy prevalent in technical analysis is indeed profitable, and if so under which circumstances. Even for a somewhat complex and extensive exercise, we will show that FRACTI models are a succinct and straightforward way to describe and communicate details and features of the trading strategy.

This example investigation will show how financial models are built by chaining "contributions" into something called streams. Everything in FRACTI is a "*contribution*"[1] and contributions are formal "*evidences*"[2] of a scientific investigation. As evidences they can be shared, reused and traced through something called a "*record of provenance*[3]".

An executable version of this exercise [10] [5] is available online and can be downloaded as a notebook [6] by any interested parties if they wish to verify the underlying data and methods. Although still a prototype and not yet intended as a complete collaboration platform, this executable shows critical features of contributions and tracking of evidences through a record of provenance that are critical for transparency, repeatable methods and computational controls in a collaborative platform.

This exercise will also demonstrate how easy it is to adjust a model in order to perform various functions in the investigation. The example of this exercise will demonstrate how slight adjustments can make the same model serve for functions of visualization, benchmarking, simulation and even real-time trading purposes if need be.

The first part of this paper is an abstract of FRACTI's blueprint presented in earlier phases of this research in previous papers. We outline the main features of the framework, main objectives and its representation system based on facets, contributions and meta-model.

The second part defines the fundaments of the scenario under investigation, a common technical trading strategy called "breakthrough crossover momentum strategy". This exercise will dissect this technical strategy in search of scenarios of profitability in two steps: first against random walks and second against historical data from components of a liquid index, the S&P 500 index.

The third part describes the end-to-end exercise of investigation using FRACTI standard process: formulation of a hypothesis; outlining of limitations, simplifications and expected outcomes; development of a model and components; definition of shocks and simulations; execution of simulations; and finally formulation of conclusions. During this exercise we generate a number of contributions – also known as evidences – and a step-by-step analysis of those evidences.

Throughout those steps we will demonstrate how easy it is to formalize FRACTI models and adjust them to cover a wide spectrum of functions, from visualization, to benchmarking, to historical simulation. The definition and execution of models generate a number of relevant shareable and traceable contributions. By the end of the investigation, using evidences produced, we list a number of possible explanations for our findings.

Finally, we conclude by listing observations about the use of the framework, evidences generated through data and visualizations and suggestions for future research.

## II. FRACTI IN A NUTSHELL

Before we jump on to define the strategy and the description of the model we will investigate it is important to describe how financials models are about to be represented and therefore we must now provide a highlight of FRACTI main aspects. This introduction is a brief outline of features extensively described in a previous paper, therefore the curious reader is strongly encouraged to refer back to the original blueprint [1] for a detailed description[4].

The definition of FRACTI was driven by very specific and straightforward objectives: to allow *transparent collaboration*, *repeatability of results*, *accessibility* and *openness*.

**Transparent Collaboration**: FRACTI defines large datasets and financial models through a common representation so that visualization and operations over large amounts of financial data is uniform. The uniform representation allows shared items to be examined in detail and re-executed against different scenarios by different groups of users. In other words, the next sections will exemplify the use of FRACTI as a **scientific support system** [11] [12].

**Reproducibility**: The conceptual framework enforces a scientific approach to analytical research: models and scenarios have to be reproducible by anyone. On this investigation exercise we will show how large sets of data and models can be traced to their origins and re-executed, allowing different organizations and individuals to easily replicate results.

**Accessibility**: End users do not have to be proficient in computer science in order to be able to use, collaborate on or visualize models or scenarios in the framework.

---

[1] "Contribution" in the same sense as authors "contribute" to Wikipedia

[2] The available body of facts or information indicating whether a belief or proposition is true or valid [65]

[3] Chronology of the ownership, custody or location of historical entities [65]

[4] Striking a balance between re-introducing concepts relevant to this paper here and repeating the content already published in the original FRACTI paper [1] is hard, so despite this brief introduction the reader is still strongly encouraged to refer to the blueprint [1] for an extensive introduction to relevant terms and concepts.

**Openness**: This investigation exercise shows instances of the trading model representing specific configurations, executions or simulations exchanged across environments or different implementations. Openness means that data and method of an investigation can be traced and replicated regardless of ownership, origin, computational implementation or location of a contribution.

The conceptual layout

t of FRACTI is in essence a proposal to **represent knowledge** in the specialized field of finance through abstractions called **models**.

The **knowledge representation system** for models is described in terms of *what* can be shared, or evidences, called in the scope of this research **contributions** and *how* to establish fundamental building blocks called **facets.**

In the scope of this work, the term **contribution** applies to artifacts[5] produced by participants (users) and shared with (or contributed to) a wider community of users.

Contributions should cover a broad range of models, methods, and results relevant to financial sciences [13]. Some examples include datasets in small, medium or large scale; time series in low, medium or high frequency; calculation methods and visualization plots; and results related to historical and real-time execution, simulation and back testing.

The term *facets*, or aspects, relate to fundamental building blocks that can be used to arrange contributions and produce more complex abstractions in a model.

The conceptual framework of FRACTI defines several different facets in order to support different cases of use in financial markets[6]. For the sake of simplicity, in the scope of this investigation exercise, we leverage one facet only: *streaming*.

The streaming aspect defines a graph-oriented Domain Specific Language [14] [15] to route fragments of meta-data $x$ through a chain of reusable and exchangeable processors $P_i$ defining a function composition [16] [17] [18]:

$$(P_1 \, o \, P_2 \, o \, ... \, o \, P_n)(x)$$

In the specific notation of the reference implementation of FRACTI in use in this investigation [5] the function composition is represented by a symbol ≫ giving a composition of processors the form:

$$x \gg P_1 \gg P_2 \gg \cdots \gg P_n$$

We refer to a chain of processors $P_1 ... P_n$ as a stream, and financial models in FRACTI are presented using a stream notation. In other contexts similar representations are called pipes, pipelines or infix notation [19].

Overall, a FRACTI representation model naturally enforces a thought process that is in line with the modern scientific method, outlined by the very specific phases:

- Formalize the phenomena under investigation. In this case we will be defining the trading strategy under investigation (Section III).
- Define a hypothesis (Section V.A).
- Define assumptions, and back-testing procedures to be followed (Section IV)
- Define components to use. Shocks, modes and benchmarks (Section IV)
- Define the expected outcome (Section IV)
- Establish conclusions (Section VI)

Despite the well-defined process outlined by these phases, they are only an indication of a best practice and not mandatory. A researcher is always encouraged to explore variations of these phases when engaging in any investigation with FRACTI.

### III. THE TRADING STRATEGY UNDER THE MICROSCOPE

The central contribution of this paper is to formalize a real end-to-end example of scientific investigation of a financial model that is self-contained, complete, simple, accessible, and widely debated in both academia and the industry: the performance of a trading strategy used commonly in technical analysis.

The concrete example described in this paper investigates a common trading strategy, a variation of a moving average cross-over (MAC-O) momentum strategy [20]. This type of hybrid strategies uses "break through" signals to identify "momentum" of a pseudo-random movement. We call this variation the Breakthrough Cross Over Momentum, or BCOM, strategy.

Breakthrough strategies are the most commonly used strategies in technical analysis and electronic trading. They can be tested with both real-time and historic data. On the correlated property, momentum strategies [21] identify profit opportunities by assuming that, unlike a purely random movement of a price (random walk), a price movement carries some "inertia" and tends to gain on an already higher price, "many more times than not … the strong get stronger and the weak get weaker" [21].

Technical analysis is usually seen as an alternative to fundamental and quantitative analysis. Technical analysis has its origins in the Dow Theory and the works of Cowles [22], Dow and Hamilton [23] [24] from early 20[th] century. Since that early time, the effectiveness of technical analysis when compared to fundamental and quantitative analysis has been a matter of controversy, with references testifying for the effectiveness of technical analysis [25] [26] and against it [27]. Different researchers test variations of these strategies using different data, which makes it very difficult to compare and contrast different researches. A much more rigorous approach is needed if we are to assess the robustness of such trading strategies.

In this exercise we will be using FRACTI models to determine the efficiency of the BCOM strategy.

We may not be able to offer a definite conclusion for this investigation vis-à-vis evidences collected. The main intent of this paper is to serve as a showcase for methods of investigation using FRACTI patterns.

---

[5] Observations in a scientific investigation or experiment that is not naturally present but occurs as a result of the preparative or investigative procedure [65]

[6] FRACTI represents financial models through the use of facets, currently streaming, reactives, distribution and simulation [1].

Models that rely on breakthrough strategies intend to predict the behaviour of a random – or pseudo-random - walk based on fluctuations of this walk crossing (or breaking through) the line given by attenuations of its past values.

In order to evaluate breakthrough momentum strategies we need to introduce each of four major components that we employed: random walks, moving averages, rules for generation of buy and sell signals and portfolio management.

### A. Random Walks

First we would like to evaluate the effectiveness of a BCOM strategy on random walk series. A common random walk is a path of successive values following some random generation pattern. The random walk theory in essence states that the path of prices over time is "efficient" and past prices have no influence over future prices [28].

One-dimension random walks are used extensively for simulation of stochastic movement of asset prices over time. In this exercise we will specifically use a Brownian motion [29], or Wiener process [30], characterized by the following properties:

$$W_0 = 0$$

$$W_t \sim N(0, t - s) + W_s$$

Where $t \to W_t$ is continuous, and $N(\mu, \sigma^2)$ is the normal distribution with expected value $\mu$ and variance $\sigma^2$ and for any $t, s$ $and$ $u$, $W_t - W_s$ and $W_u$ are independent for $u \leq s < t$.

In FRACTI parlance we can illustrate a random walk using a straightforward model of just one stream in one line[7]:

```
ts('2013-01-01', '2013-12-31') \
>> brownian(seed=42, s0=37) \
>> plot
```

**Figure 1. Random Walk Over a Time Series**

On this dialect we describe one stream connecting a daily time-series from beginning to end of 2013, a Brownian motion random walk[8] for $W_0 = 10$. The resulting sample is visualized by a bi-dimensional plot, as shown in Figure 2:

---

[7] We emphasize that we have not implemented a full system supporting concepts that we present in this document. This specific dialect is provided for illustration purposes only.

[8] On all instances of this exercise "seed" arguments are provided in order to achieve repeatability of results of a random series

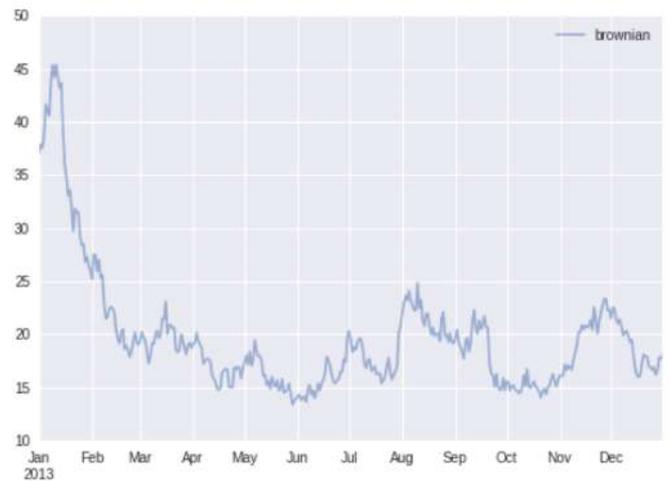

**Figure 2. Single One-Dimension Brownian Motion**

For simulation purposes, we can generate several Brownian time series. For example, we can generate in one line series of prices mimicking random walks for closing prices of three symbols, GOOG, IBM and B:

```
q.seed(42)
ts('2013-01-01', '2013-12-31') \
>> brownian(s0=37, output='GOOG') \
>> brownian(s0=21, output='IBM') \
>> brownian(s0=42, output='B') \
>> plot
```

**Figure 3. Multiple Random Walks Over a Time Series**

This single line defines a stream connecting to a daily time series for the year 2013, three separate random walks with different $W_0$ for GOOG, IBM and B. As before, all results aggregated in a bi-dimensional plot:

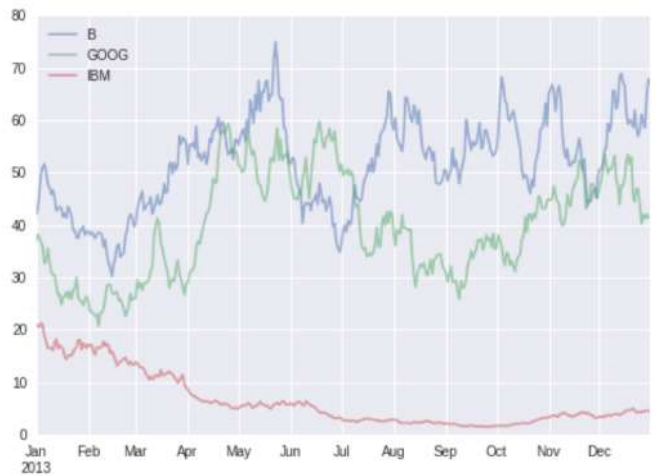

**Figure 4. Multiple One-Dimension Brownian Motions**

### B. Moving Averages

The second major component in evaluating a BCOM strategy is the generation of an attenuation signal. Filtering, or dampening, of the random walk is achieved by running or rolling averages[9] of past prices in a time series updated on each new tick of the price.

---

[9] Arithmetic mean

Moving averages in their formal form are non-recursive, and by consequence computing intensive [31]. For every new price, all past prices would have to be traversed and a new average calculated. Part of the exercise depicted in this paper represents calculations in a recursive form, in which iterations over past values of a series are not necessary. Whenever available, we introduce a model's non-recursive form, to provide a means of comparison to models available in other references [32], before showing the recursive form.

To thoroughly evaluate the BCOM strategy, we use different moving average calculations. The most common types of moving averages are cumulative, rolling, weighted and exponentially weighted moving averages [32].

**Cumulative Moving Average**: The simplest moving average is simply an arithmetic average of the previous $n$ values in a series. The non-recursive and recursive calculation are given respectively by the models:

$$CMA_n = \frac{1}{n} \sum_{i=0}^{n} x_i \quad \text{(non-recursive)}$$

$$CMA_n = \frac{1}{n} (x_n + (n-1) CMA_{n-1}) \quad \text{(recursive)}$$

This is a special case of moving average where there is no sampling window, so all data is considered equally in the calculation of the moving average.

**Rolling Moving Average**: The "rolling" average is an unweighted mean of previous samples, considering a sampling window of size $m'$. A more precise definition accounting for $n$ where $n < m'$ the window is given by $m = \min(m', n)$. The following models give the non-recursive and recursive forms:

$$RMA_n = \frac{1}{m} \sum_{i=(n-m+1)}^{n} x_i \quad \text{(non-recursive)}$$

$$RMA_n = RMA_{n-1} + \frac{1}{m}(x_n - x_{n-m+1}) \quad \text{(recursive)}$$

**Weighted Moving Average**: Weighted averages have a dampening factor ($m$) assigning different weights to data at different positions in a movable sample window of size $m'$. A more precise definition accounting for $n$ where $n < m'$ the window is given by $m = \min(m', n)$.

In this moving average $m$ past factors are adjusted by a decreasing linear factor $(m - n + 1)$, so that more recent observations have a larger influence on the filtered signal. Its form is given by:

$$WMA_n = \frac{m.x_n + (m-1).x_{n-1} + \cdots + x_{n-m+1}}{m + (m-1) + \cdots + 1}$$

Where the denominator $\sum_{i=0}^{m-1}(m-i)$ is a triangular number [33] that can be reduced to $\frac{m(m+1)}{2}$, giving the model a final form:

$$WMA_n = \frac{2}{m(m+1)} \sum_{i=1}^{m} i \cdot x_{n-m+i}$$

This model is partially recursive from 0 to the value of $m$ for each iteration $n$.

**Exponentially Weighted Moving Average**: In the exponentially weighted moving average a dampening factor $\alpha$ is used to decay older terms of the series. Its recursive form is given by:

$$EWMA_n = \alpha x_n + (1 - \alpha) EWMA_{n-1}$$

Where $0 < \alpha < 1$. Higher dampening factors ($\alpha$) would give lower weight to older terms of the series, yielding slower filters. This model is recursive by definition and therefore easily adapted to computational forms.

We can bind together multiple filters to a random walk and to a visualization endpoint on the same stream using one single line.

```
ts('2014-1-1', '2014-12-31') \
>> brownian(seed=42, s0=37) \
>> ewma(alpha=0.2) \
>> rma(m=20) \
>> cma \
>> plot
```

This stream defines a random walk with $W_0 = 10$ over a time series for the year 2014, adding three different filters: EWMA (exponentially moving average) with $\alpha = 0.2$, RMA (rolling moving average) with $m = 20$ and CMA (cumulative moving average) chaining it all for the following visualization of results:

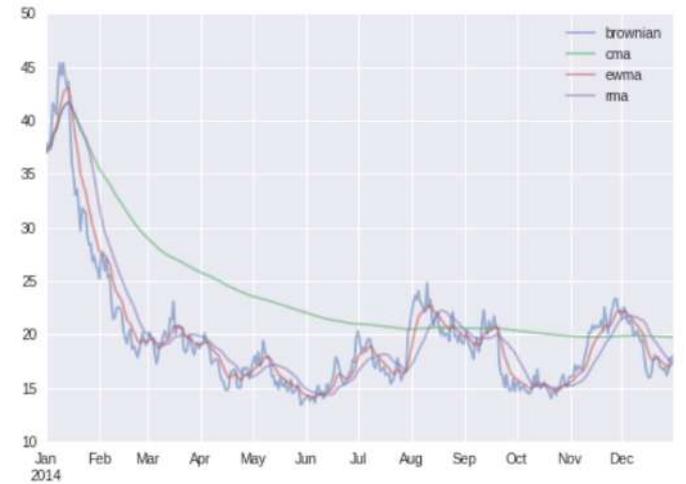

**Figure 5. Multiple Filters over a Random Walk**

Momentum strategies tend to rely on filters that provide some control over how much decay should be applied to older terms of the series. Additionally, as explained before, best results are achieved from filters that can be represented recursively. Given these advantages we will be using the exponentially weighted filter in this investigation.

### C. Derivation of Buy and Sell Signals

The third major component for evaluating a BCOM strategy is the generation of buy and sell signals. In a BCOM strategy buy and sell signals $S_i$ are generated depending on how a "fast" curve of prices $F_i$ crosses over a "slow", or dampened, curve of prices $D_i$.

In essence, a buy signal is generated whenever the fast curve crosses the slow curve upward. On the other hand, a sell signal is generated whenever the fast curve crosses the slow curve downward, i.e.:

$$S_{t+1} \rightarrow SELL \quad if \ D_{t-1} > F_t \ and \ D_{t-2} \leq F_{t-1}$$
$$S_{t+1} \rightarrow BUY \quad if \ D_{t-1} < F_t \ and \ D_{t-2} \geq F_{t-1}$$

Where $(t-2, t-1, t)$ denotes the three steps in the time sequence that are relevant to determine if the signal $S$ in the immediately subsequent step $t+1$ will be either a sell or a buy signal.

A clear illustration of this behaviour is given in Figure 7. In that plot we denote $SELL$ signals by red triangles whenever a cross happens on D over F ($D_{t-1} > F_t \ and \ D_{t-2} \leq F_{t-1}$) and $BUY$ signals by green triangles whenever a cross happens on F over D ($D_{t-1} < F_t \ and \ D_{t-2} \geq F_{t-1}$).

We define this strategy as a $maco$ (moving average cross over) function, which takes a time series as input and outputs a set of buy and sell signals.

In this model there are basically two variations of rules for the derivation of signals, depending on what makes the faster and slower curves. These variations are called double and single cross over rules. In a double cross over variation both the slower and faster curves are themselves filtered curves of a spot price curve. The slower curve has a higher dampening factor ($\alpha$) than the faster curve.

If we were to use moving averages as dampening filters, for example, the fast curve $MA''_i$ and slow curve $MA'_i$ would basically differ by how far back in the past spot prices they used in the average calculation: in other words, slow moving averages track longer periods than fast moving averages does, i.e.:

$$F_i \approx MA''_i$$
$$D_i \approx MA'_i$$

In a single cross over variation the faster curve is given by the spot price over time $F_i$ and the slower curve is given by the dampening of this spot price through some filter, usually a moving average $MA_i$, i.e.:

$$F_i \approx W_i$$
$$D_i \approx MA_i$$

In this exercise we will be using a single cross over variation, in which the model takes the final form:

$$S_{t+1} \rightarrow SELL \quad if \ MA_{t-1} > W_t \ and \ MA_{t-2} \leq W_{t-1}$$
$$S_{t+1} \rightarrow BUY \quad if \ MA_{t-1} < W_t \ and \ MA_{t-2} \geq W_{t-1}$$

Where $W_i, MA_i, S_i$ are respectively values of the stochastic random walk, moving average filter and signal buy or sell at time $t = i$.

### D. Portfolio Management

The fourth and final step of a trading model should account for portfolio management by keeping track of overall gains and losses, taking into consideration the cash flow resulting from buys and sell signs, transaction costs, a cash balance, and price variations of the underlying:

$$(P, B, \alpha)_t = pm(K_t, S_t, L_t, W_t)$$

The function $pm$ is the portfolio management function where the arguments $K_i, S_i, L_i, W_i$ are respectively the cash balance, signal (Buy or Sell), load (transaction costs) and the value of the stochastic random walk at time $t = i$. The function pm takes buy and sell signals generated by the crossover momentum strategy as input. Based on the cash balance ($K$), pm outputs a 3-tuple $(P, B, \alpha)_t$ of the model at time $t$.

In later references in this paper, the notation for portfolio management is equivalent to:

$$pm \approx cash\_stock$$

In this tuple $P$ is the final signal to be sent to the market place (Buy, Sell or Nothing); the full portfolio balance $B$ accounting for the summation of cash and non-cash positions; and the overall profitability of model denoted by $\alpha$.

## IV. THE FRACTI REPRESENTATION

We have explained the foundations of FRACTI and the BCOM trading strategy under investigation. We can now move ahead and describe the representation of BCOM using FRACTI concepts.

As described in Section II, all financial models in FRACTI are represented through steps in a sequence of a particular type of facet, a stream. Despite the relative complexity of the model described in Section III, we can outline the entire strategy to evaluate BCOM by the following steps:

- Generate price ticks, either from random Brownian generators or from historical data;
- Generate moving average, preferably allowing plugging in different types of moving averages;
- Generate buy or sell signals based on breaks of the spot price curve through the moving average;
- Decide whether to send an order to the market or not, based on current portfolio and liquidity (balance of available cash);
- Processing the resulting data graph (plotting, store and track for future use, etc.).

We can generate in one line an example of these steps. A model of a daily time-series for years 2013 and 2014, a simulation of stochastic random walk of closing prices, a generation of buy and sell signals based on cross overs, portfolio management and visualization is given by the following description:

```
ts('2013-1-1', '2014-12-31') \
>> brownian(seed=42, s0=37) \
>> ewma(alpha=0.05) \
>> maco \
>> cash_stock(initial_cash=10000, load=7.5) \
>> plot
```

**Figure 6. Breakthrough Momentum Strategy Model**

This brief representation of this one line stream is the full model for a time-series for years 2013 and 2014 ($ts$); a simulation of a stochastic random walk of closing prices ($brownian$) with $W_0 = 37$; an exponentially weighted filter ($ewma$) with $\alpha = 0.05$; a moving average cross over ($maco$) and a portfolio management function $pm$ ($cash\_stock$)) with $K_0 = \$10,000$ and $L_i = \$7.5$.

The resulting contribution of the execution of this stream is a visualization plot, the first quick and structured glimpse into what to expect from a BCOM strategy, shown in Figure 7.

In blue we see the random walk simulating closing prices of an underlying hypothesis. In green we see the EWMA filter. The green and red triangles show when buy and sell signals are sent to the market. Finally in red, is the profitability over time (on the right vertical axis, in percentage points) of the overall strategy, given the features discussed before.

A profitable strategy would show alpha (red line) greater than one, so our own definition of a profitable strategy would be:

$$\alpha > 1.0$$

In a nutshell, this first run simulates a random walk, with the specific parameters of $W_0 = 37$, $EWMA_\alpha = 0.05$, $K_0 = \$10,000$ and $L_i = \$7.5$ in which a BCOM strategy is not profitable, losing about ~6.5% overall in two years.

Models are in essence simulated simplifications of a real world phenomenon [34] [35]. This BCOM model is no exception. In order to allow proper computational representation, this FRACTI representation of BCOM will introduce a few important simplifications.

- Support for one single order book, in other words, one symbol of the underlying asset. The conclusions should be assumed for additional symbols following the same price behaviour.

- Infinite market liquidity. The marketplace guarantees market orders to be fully executed over the cycle of the next price tick.

- No price lagging. The marketplace guarantees market orders to be executed on the last price tick received.

These adjustments help with the explanation, but should not substantially impact the generality of the experiment under investigation.

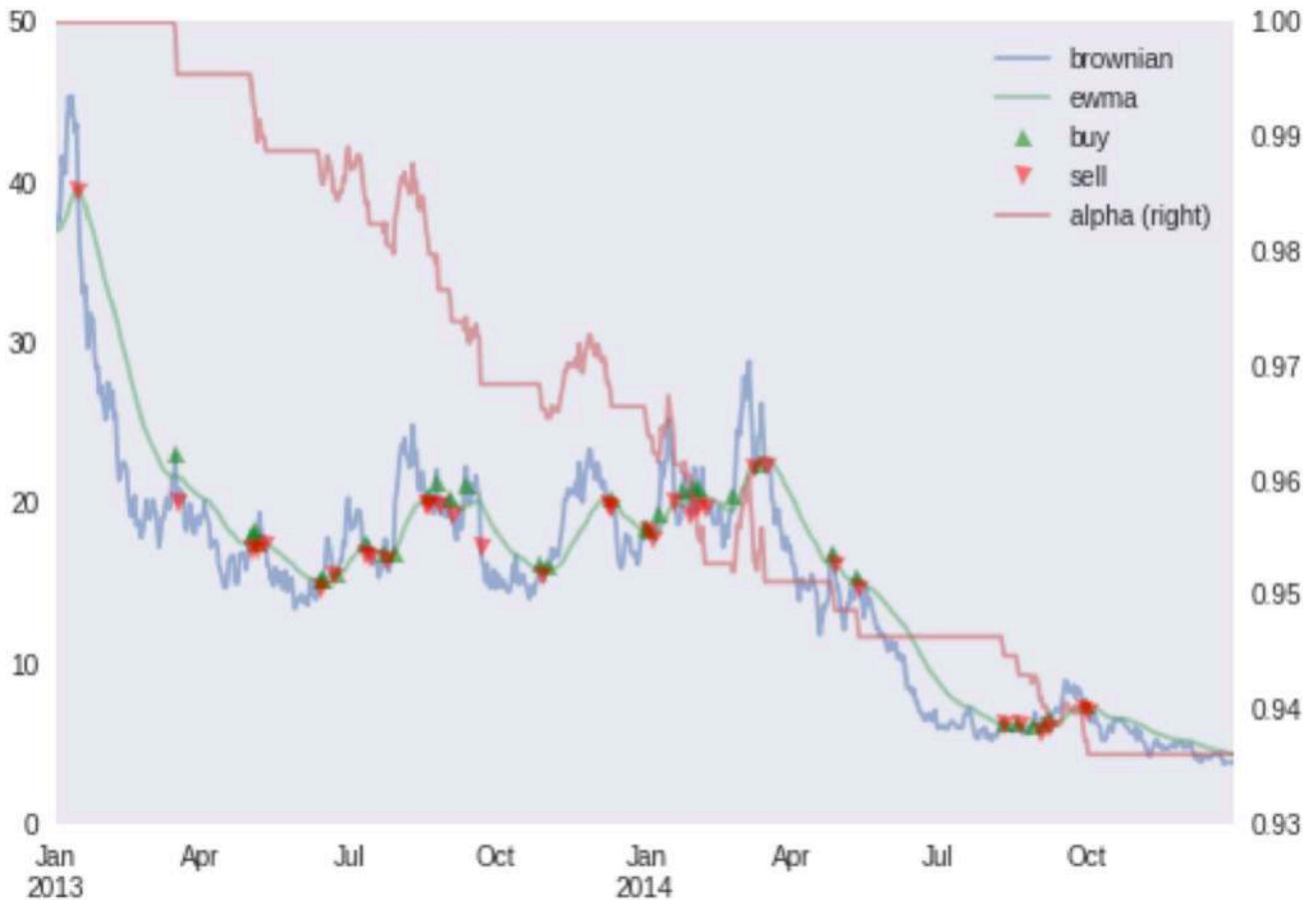

**Figure 7. Visualization of a Breakthrough Momentum Strategy**

V. THE INVESTIGATION EXERCISE

By now we have all we need to start the investigation exercise. We have explained the foundations of FRACTI and the BCOM strategy is understood and defined in terms of FRACTI concepts. We have built a very good idea of what we intend to measure and, as a consequence, are able to prove or disprove. We will now move forward with the investigation per se. Steps of the procedure described here are available as a FRACTI scratchpad (notebook) [6] and can be inspected online [10].

Given first brief results from Figure 7 there are a few immediate questions that need to be addressed and these will form the base of the remainder of this exercise:

- Are breakthrough momentum strategies money losers? Or are they ever profitable?
- If they are profitable, what features, if any, do we need to fine tune in order to make them consistently profitable?

Considering these preliminary inquiries, the first step on the scientific investigation method is to state our hypothesis.

A. *The Hypothesis*

We hypothesise the following for this exercise:

- There are scenarios under which momentum strategies are consistently profitable.
- For some combinations of $W_0$, $EWMA_\alpha$, $K_0$ and $L_i$ we expect the momentum strategy to be consistently profitable.
- If profitable against a random walk, we expect the strategy to be profitable against a representative sample of financial instruments that follow a quasi-stochastic price movement path.

We will test the above hypotheses on two major cases: first on a Monte-Carlo simulation using stochastic generators on variations of arguments of $W_0$, $EWMA_\alpha$, $K_0$ and $L_i$; and second, back testing against constituents of a well-known index, the S&P 500 index [36].

B. *Monte Carlo Simulation of Brownian Variations*

The first part of this exercise is an attempt to examine the first portion of our hypothesis: finding out for which combinations of parameters of a random walk $W_0$, $EWMA_\alpha$ and $L_i$ we should expect the momentum strategy to be consistently profitable.

We will attempt to answer that by defining a model that executes on different shocks. The term "shock" in FRACTI nomenclature denotes "one single iteration" of a simulation model. In other words, each shock carries one permutation of values of features relevant for a specific simulation.

In this exercise, each shock carries a variation of features: initial value of the random walk $shock.s0$ ($W_0$); the dampness factor of the filter $shock.alpha$ ($EWMA_\alpha$) and transaction costs $shock.load$ ($L_i$).

```
def momentum_simulation(shock):
    return ts('2013-1-1', '2014-12-31') \
    >> brownian(s0=shock.s0) \
    >> ewma(alpha=shock.alpha) \
    >> maco \
    >> cash_stock(initial_cash=10000, load=shock.load)
```

**Figure 8. Simulation Model**

The result of the execution of the model defined in Figure 8 is the feature $fitness$ ($\alpha$), or overall profitability of the model. This feature is displayed on both x and y axis of the resulting plot in Figure 9.

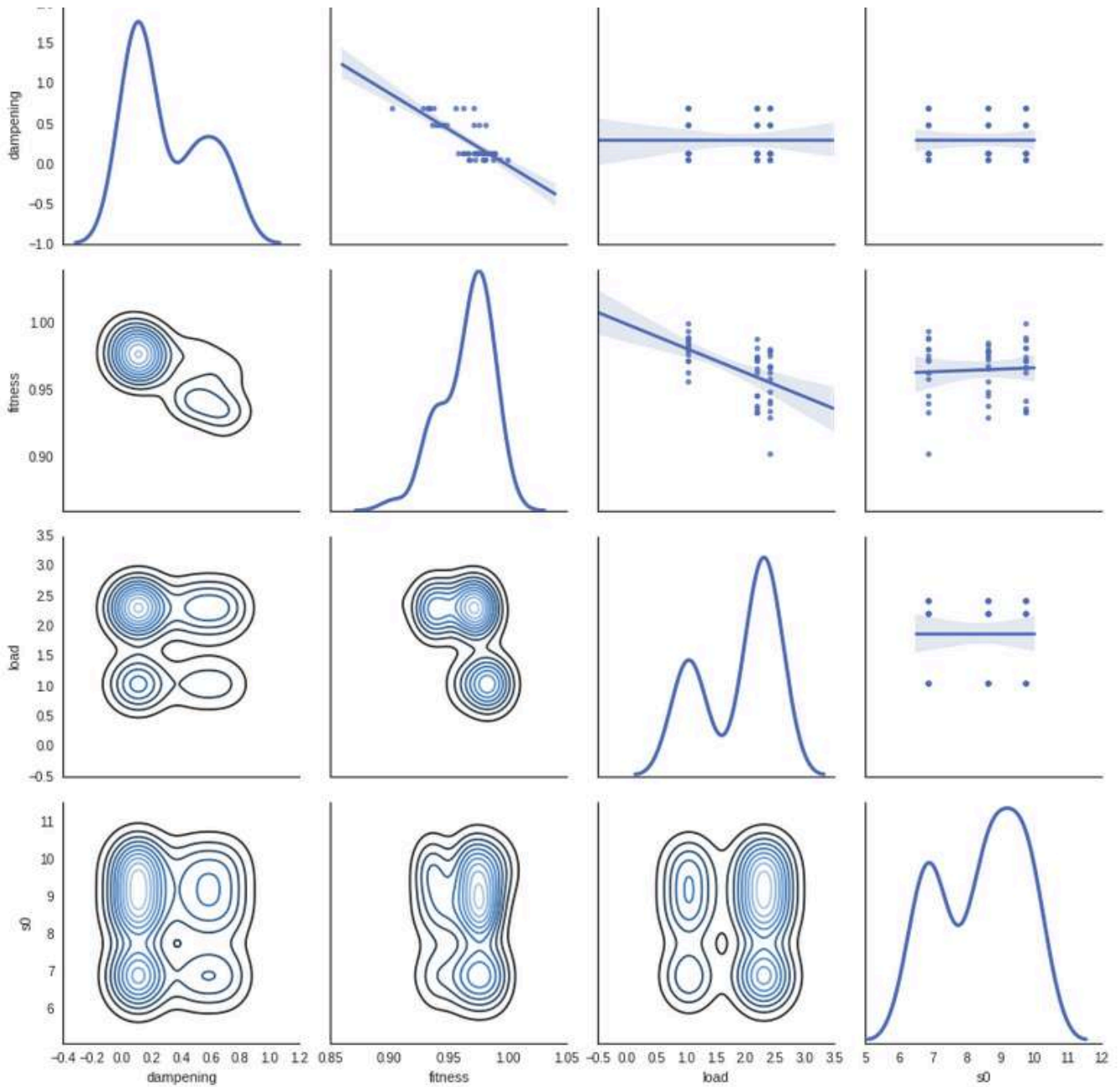

**Figure 9. First Monte Carlo Simulation, Scatter Plot Matrix**

In this investigation exercise, for illustration purposes only, we intend to investigate correlations using linear regressions and scatter plot matrices[10] [37].

The diagonal of the scatter plot brings a normal interpolation of each of the features, from where we can visually get a sense of mean and distribution for each of the features. The upper half of the matrix brings a scatter plot of each pair of features, along with a linear interpolation of the pair. This gives us a sense of mutual correlation. The lower half of the matrix shows a cluster plot where we can observe any patterns of clustering on each of the pairs.

If one compares Figure 6 to Figure 8 one can see that they essentially describe the same model. One could use the same stream to switch from visualization to a Monte Carlo simulation on the same model immediately, just changing from constants to arguments in a shock.

The next step, after defining the composition of a shock, is to perform a simulation. In this exercise, for illustration purposes we decided to use uniform samples to represent variations on each of the arguments.

In a uniform distribution, if $rs(n)$ is a random sample of size $n$, the continuous uniform distribution in the range $[a, b]$, denoted by $Unif[a, b]$ for $b > a$, is given by:

$$Unif(a, b) = (b - a) * rs(n) + a$$

In this model the continuous uniform distribution using the same arguments is defined by $unif(a, b, n)$.

---

[10] Usually real-world investigation scenarios will rely on a much higher number of features, which would make the use of visual methods impractical. Numerical methods would be more appropriate.

We "shock" the model defined in Figure 8 with values in $shock.s0$, $shock.alpha$ and $shock.load$ as permutations of values in uniform distributions. The entire definition of the procedure is done in one line:

```
b = montecarlo(momentum_simulation, \
               s0=unif(5.0, 10.0, 3),\
               alpha=unif(0.01, 0.8, 5),\
               load=unif(1.0, 3.0, 3))
```

This executes our model against permutations of each of the features, given random values taken from a uniform distribution $unif(a, b, n)$, and generates Figure 9.

From top to bottom and left to right a scatter plot matrix, like the one presented in Figure 9, presents the same sequence of features in horizontal and vertical.

In this case these are in order $EWMA_\alpha$, $\alpha$, $L_i$ and $W_0$ – respectively representing dampening of the EWMA, fitness (or profitability), load (or transaction costs) and initial price of the instrument under simulation.

For the first try of the investigation, we generate the scatter plot matrix in Figure 1 using one statement:

```
b >> scatter_matrix(index=None)
```

The diagonal, upper and lower halves of the matrix give us a deeper insight on features under study and the behaviour of the model overall:

- This strategy is still never profitable against random walks using the range of uniformly distributed arguments for this simulation. No shocks were able to bring $\alpha > 1$, our own definition of "profitable" as explained in Figure 7.

- The lower the dampening of the filter, the less money is lost. This is shown by the (dampening x fitness) scatterplot on the right side half of the scatter matrix, row 1 and column 2, with a negative line-of-fit. In other words, the strategy will lose less money using slow filters; or using yet another phrasing, $EWMA_\alpha$ (dampening) is negatively correlated to $\alpha$ (fitness).

- As one could expect, this simulation shows that the lower the transaction costs (load), the less money is lost. The lower the feature $L_i$ (transaction costs, or load), the higher $\alpha$ (profitability, or fitness).

- As one could intuitively expect, the initial price of a stock has no influence on the profitability of this model.

This is shown by the (s0 x fitness) scatterplot on the right side half of the scatter matrix, with a virtually neutral line-of-fit. Feature $W_0$ has no correlation to $\alpha$.

We can see that one of the features is irrelevant for what we are investigating. The quick inspection described above showed $W_0$ (initial price of a stock) has no influence over $\alpha$ (profitability or fitness) and should be removed. On that note, we will adjust our model to remove $W_0$ and add a new feature $W_\sigma$, namely the variance of the Brownian random walk.

```
def momentum_simulation_modified(shock):
    return ts('2013-1-1', '2014-12-31') \
    >> brownian(sigma=shock.sigma) \
    >> ewma(alpha=shock.alpha) \
    >> maco \
    >> cash_stock(initial_cash=10000, load=shock.load)
```

Our new model basically represents the same as Figure 6 and Figure 8. However, it functionally does something substantially different: this new model investigates if the variance of a random walk ($W_\sigma$) affects profitability ($\alpha$), and if so under what circumstances.

The new shock features top to bottom and left to right are, in order, $EWMA_\alpha$, $\alpha$, $L_i$ and $W_\sigma$ – respectively representing dampening of the EWMA, fitness (or profitability), load (or transaction costs) and variance of the Brownian motion. A scatter plot of this simulation is given in Figure 1.

We can see that all findings on the first try still hold true for the second try, with an extra insight. Now we get an additional conclusion about the correlation between $W_\sigma$ and $\alpha$: the (sigma x fitness) cell on the right side of the scatter matrix shows a scatterplot with a negative line-of-fit. This indicates that the variance of a random walk (sigma, or $W_\sigma$) is negatively correlated to profitability (fitness, or $\alpha$) or, in other words, we should expect to lose slightly less money when a random walk presents a lower volatility.

The second try also confirms the bottom line of this simulation: against a random walk this model is not profitable.

### C. Back-Testing Against the S&P 500 Index

For the second part of our original hypothesis we will back-test the model for profitability against historical price data. For the sake of transparency we selected to use constituents of the S&P 500 index. In essence this index, created in 1957, tracks US stocks with at least USD 5.3 billions of market cap [38].

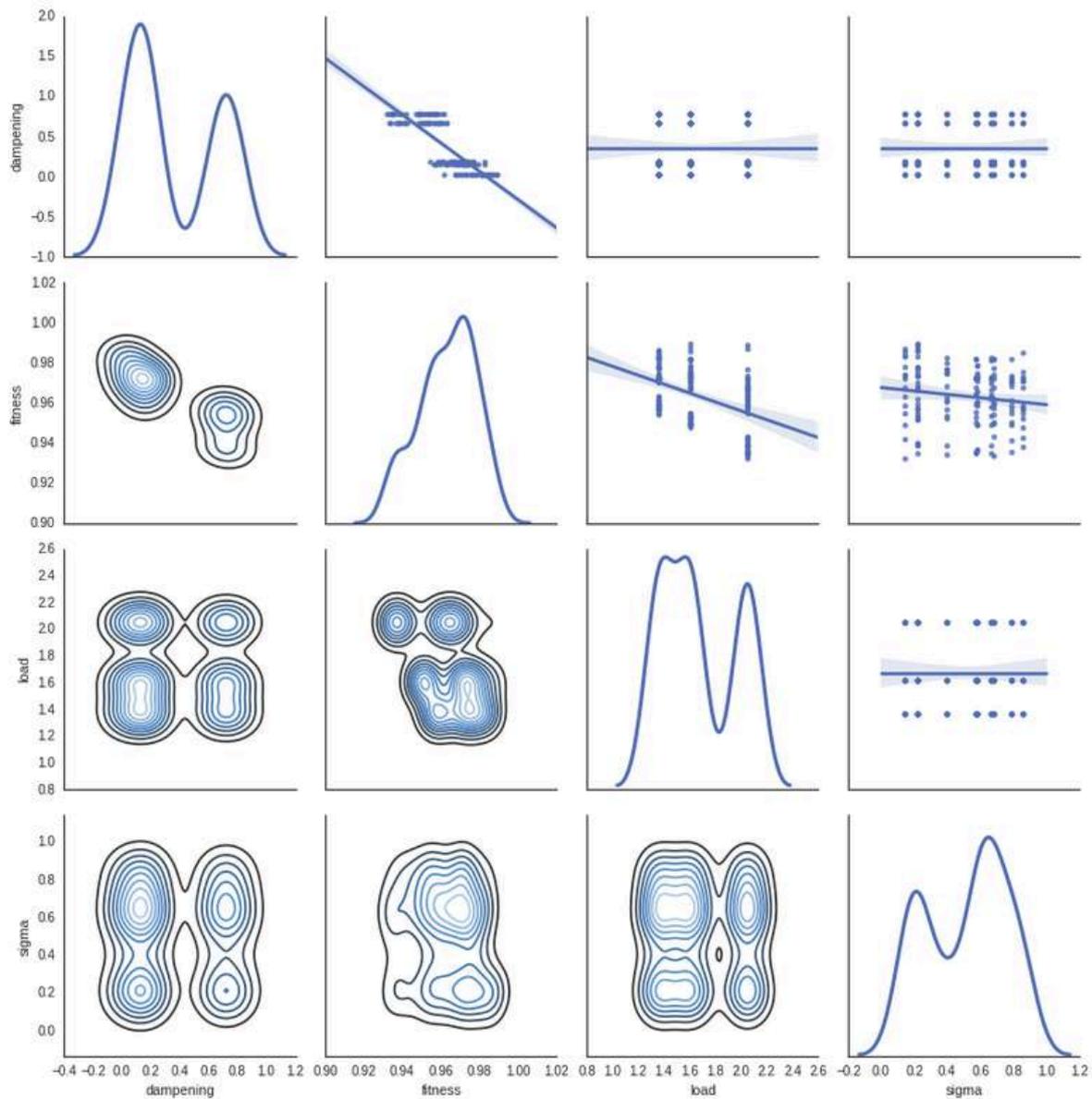

**Figure 1. Second Monte Carlo Simulation. Scatter Plot Matrix**

Historical data is also a contribution in FRACTI terms, and as such it can be leveraged as part of streams and be bound to other contributions. As an example, we can briefly inspect one stream of historical data in one line:

```
historical('AAPL', '2014-01-01', '2014-12-31', columns=['Adj. Close']) \
>> plot(index='Date')
```

This stream produces a plot contribution of adjusted closing prices[11] for Apple Computers for the year 2014:

---

[11] A stock's closing price on any given day of trading that has been amended to account for distributions and corporate actions that occurred at any time prior to that day's closing [66].

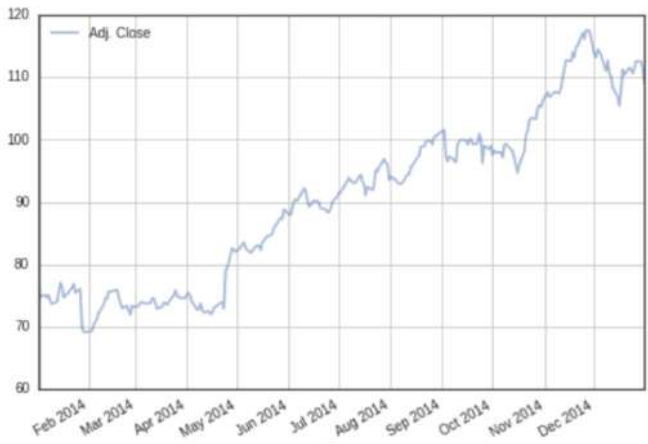

**Figure 10. Adjusted Closing Prices for AAPL in 2014**

We can leverage historical data on our model like any other contribution. If we want to use historical data instead of a random walk in our model, all we have to do is replace the first and second steps, $ts$ and $bownian$, with one step: $historical$:

```
historical('AAPL', '2014-01-01', '2014-12-31', columns=['Adj. Close']) \
>> ewma \
>> maco \
>> cash_stock(initial_cash=10000, load=7.5) \
>> plot
```

This slight modification leverages basically the same original model, this time to create a plot contribution for closing prices of Apple Computers for the year 2014 in Figure 11 :

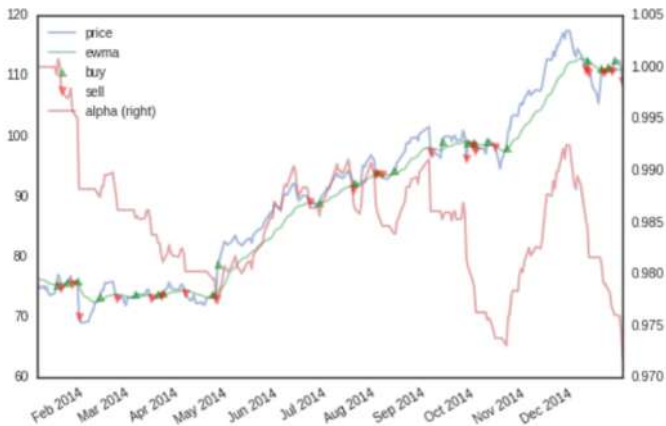

**Figure 11. Historical Performance of AAPL**

The original scenario seems to be repeating itself for historical data. If one were to start using this strategy to trade APPL stock at the beginning of 2014 one should expect to lose approximately 3% of the original cash balance. In other words, of the initial $10,000 investment in APPL stocks on 1/1/2014, only about $9,700 would remain by 12/31/2014.

Is this something specific to Apple Computers stock? How does a different stock, like Google (GOOG), behave in the same period using the same strategy features?

```
historical('GOOG', '2014-01-01', '2014-12-31', columns=['Adj. Close']) \
>> ewma \
>> maco \
>> cash_stock(initial_cash=10000, load=7.5) \
>> plot(out='goog_momentum.png')
```

To answer this question, only one slight modification – a change of a constant from 'APPL' to 'GOOG' - is required in the proposed framework to visualize the historical performance of a different instrument:

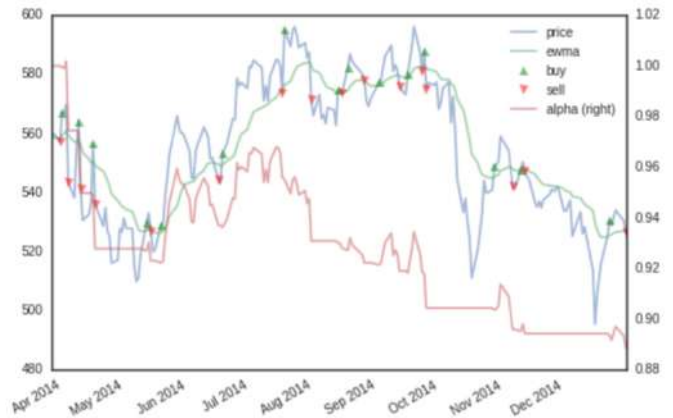

**Figure 12. Historical Performance of GOOG**

Considering a new security on the simulation, Google, we lost even more; 12% of the original investment was lost over the same period.

Still, we have run the model twice, on two stocks, and neither was profitable. One could argue that the sample is not representative of an expected behaviour, and both are in the same sector and both are large cap stocks. The bad performance we saw in both exercises could be related to poor data selection and bias.

To investigate further, we could back-test this model against all stocks constituents of the S&P 500 index. This should be representative of stocks of varied sectors and reasonably large caps.

As we did on the first exercise, the first step is to define the simulation model:

```
def snp500_model(shock):
    stream = historical(shock.symbol, '2014-01-01', '2014-12-31', columns=[shock.column]) \
    >> ewma \
    >> maco \
    >> cash_stock(initial_cash=10000, load=7.5)
```

Again, under our framework, this only requires a slight modification of the model in Figure 6, except that in this case only two features are necessary:

- The symbol of the stock
- The feature, or column, on the historical repository

We retrieve all constituents of the S&P 500 index and execute the benchmark of all adjusted close prices ('Adj. Close') for all symbols in one statement:

```
tickers = index('SP500') >> select(lambda x: x['Ticker'])
benchmark(snp500_model, symbol=tickers, column=['Adj. Close']) \
>> select(lambda x: x.fitness) \
>> hist
```

This benchmark runs in two steps:

- We parse the index repository 'SP500' for all symbols ('Ticker') of S&P 500 constituents

- We benchmark each shock, in which each shock has a symbol and a column. We collect the fitness of the model for this specific symbol, and plot a histogram with the distributions of results.

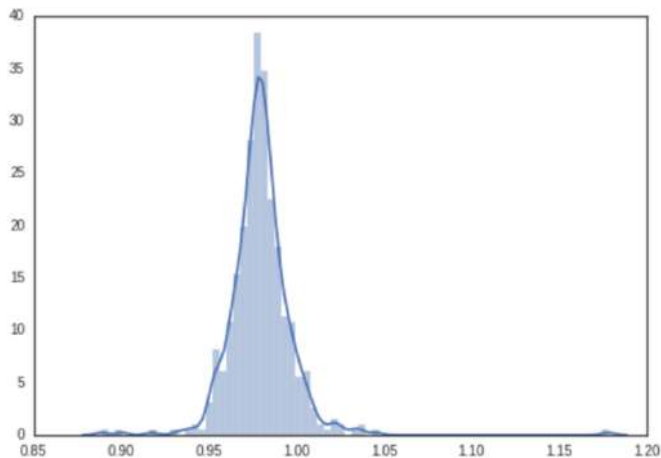

**Figure 13. Distribution of Results, Simulation S&P 500**

The result is given by a histogram fitted by a normal approximation, given in Figure 13. The results closely fit a normal distribution, and the bottom line of this part of the exercise is clearly demonstrated:

- Evidences collected here indicate that this strategy is not profitable for stocks in the S&P 500 index.
- There is one single outlier in which we can achieve a return of approximately 18%.
- On average, a stock in the S&P 500 index picked randomly and traded with a BCOM strategy would be losing around 2% for the year 2014.

### D. Provenance of Contributions

Everything in FRACTI is a contribution, and as such they all carry amongst other things a record of provenance. To illustrate that, we investigate in this section the record of provenance of Figure 12, a performance plot generated previously in this exercise.

```
provenance('goog_momentum.png')
```

This statement introspects the record of provenance of Figure 12, short-named on generation as 'goog_momentum.png', showing all details related to ownership, time stamps, versioning, source and transformation steps for all data:

```
**** PROVENANCE hdf://quantlet/jfaleiro/goog_momentum.png ****

quandl.get('https://www.quandl.com/data/WIKI/GOOG')
`-- v N/A jfaleiro @ 8/15/2015 14:35:12 EST

historical.cache('hdf://quantlet/pub/quandl/WIKI/GOOG.h5')
`-- v 0.0.1 jfaleiro @ 8/15/2015 14:35:12 EST MD5:d9d914b9bdcd9fccd913d4ab77909dbf

stream
    quantlet_analytics.dataset
        historical('GOOG', '2014-01-01', '2014-12-31', columns=['Adj. Close'])
        `--- v 0.0.1 jfaleiro @ 1/1/2015 8:34:02 EST MD5:29bc31efd050df78976a2b85250c2e54
    quantlet.filter
        ewma
        `--- v 0.0.1 jfaleiro @ 1/1/2015 8:34:02 EST MD5:9ed435ab2b00f1afdf4ff79cc74e4424
    quantlet.strats.momentum
        maco
        `--- v 0.0.1 jfaleiro @ 1/1/2015 8:34:02 EST MD5:9ed435ab2b00f1afdf4ff79cc74e4424
    quantlet.strats.portfolio
        cash_stock(initial_cash=10000, load=7.5)
        `--- v 0.0.1 jfaleiro @ 1/1/2015 8:34:02 EST MD5:9ed435ab2b00f1afdf4ff79cc74e4424
    quantlet_analytics.plot
        plot(out='goog_momentum.png')
        `--- v 0.0.1 jfaleiro @ 9/1/2015 8:15:35 EST MD5:ecfc5d3f498074e79cbb0b309dc1b786

file('hdf://quantlet/jfaleiro/goog_momentum.png')
`-- v N/A jfaleiro @ 9/17/2015 21:05:02 EST MD5:72802afc4283d8077da1cb368aa6c2cd

**** end of hdf://quantlet/jfaleiro/goog_momentum.png ****
```

**Figure 14. Record of Provenance of GOOG Plot**

There are a number of important details we can infer from a record of provenance:

- The source of all data and steps followed for generation are public and transparent. Ownership of each contribution (user jfaleiro), timestamp of each contribution (date and timestamp at which the contribution was generated) and source (indicated by an URI) are shown explicitly.
- The data used to create this plot was obtained on-line [39] on a specific date shown on the record, and was passed through a number of stages – the stages themselves being contributions – for caching and transformation, according to a specific stream. The generation of this version of the plot is also stated on the record of provenance.
- There is an URL associated to any contribution (on the case of this plot contribution the URL is given by hdf://quantlet/jfaleiro/goog_momentum.png:0), indicating location and version. This URL allows this contribution to be shared with any collaborator with knowledge of this URL and proper credentials.
- All contributions are signed, safe stored, and check summed.

Provenance tracking is one of the fundamental features of FRACTI that allows for sharing, tracking and collaboration among heterogeneous parties through which we achieve reproducibility in large-scale scientific research [1].

### E. Final Notes on Evidences of Profitability

There are two extreme opposing views regarding the efficiency of technical strategies. They vary from derisively equating technical strategies to "tea leaves reading" [40], "black magic" [41] or "financial astrology" [42] [43], to several claims on the other end of the spectrum as this model's consistent return in the range of double-digit percentage points in a year. [44]

Mimicking those same opposing views, but avoiding extreme stances, peer-reviewed scientific publications are apparently divided between supporting and repudiating claims of efficiencies of technical strategies [45] [46] [47]. A survey of past studies indicated that from "95 modern studies, 56 concluded that technical analysis had positive results", and pointed to the difficulty in getting to conclusive findings due to "data-snooping bias and other problems" [45] and "noise in trading price" [48].

Adding to these previous findings, the first simulation used on this paper observed that fully random walks were consistently unprofitable. On the second simulation, using real historical data shows that this model was only profitable in about 8% of cases.

We were not able to achieve claimed results indicating a consistent profitable behaviour using either random walks or real historical data. Evidences produced in this exercise suggest that the strategy is not consistently profitable and the hypothesis we had outlined at the beginning of this exercise is false.

These findings show strong discrepancies from claims from several investment resources. The specific causality is difficult to assess, but we can list possible explanations:

1. Data snooping[12] [49] and survivorship biases[13] [50]. Evidences in this paper might be getting the same results as previous studies did, in which technical "rules are profitable when considered in isolation, but these profits are not statistically significant after" adjustment for data snooping and survivorship bias [27].

2. Our data samples are too efficient. The very notion of a pure random walk contradicts the assumptions followed by chartists that price movement carries some "past memory" or private information [28]. In that sense, random walks and components of the S&P 500 Index might be just too efficient for momentum strategies to perform. "There is some evidence that technical trading rules perform better in emerging markets" due to their inefficiencies [27], technical strategies tend to perform better in inefficient markets [51].

3. Traders might be using variations of this momentum strategy that are actually profitable, and not disclosing its details.

4. Either data or algorithms we relied on for these calculations are wrong or have bugs. Despite care, multiple reviews and regressions over this model, inaccuracies of this kind are unfortunately commonplace in scientific research [52] [53] [54] [55] [56] [57]. In contrast, all contributions in this exercise are available, traceable and verifiable by any interested parties if needed. This transparency of evidences is indeed the main motivation behind FRACTI [1].

To derive causality from computational artifacts that seem correlated at first sight is a hard task, especially when scenarios are not exhaustive. As we have discussed in previous papers [1], we should expect this determination to become increasingly more difficult as we have to deal with higher volumes of data and more complex representations, the consequence of the "informatics crisis" [12] and the noise present in scientific investigation [1].

As it is clear by now, in this paper the determination of the exact cause for a phenomenon is secondary. The primary objective is to demonstrate how simple it is to adjust and modify models and simulations in order to follow fluid ideas and how the conceptual framework enforces collaboration through shared evidences (FRACTI contributions). We cannot determine the cause of discrepancy, and the answer will remain debatable and possibly the subject of future research. We argue that extensive and robust analysis can only be supported by collaborative research. FRACTI is a platform that supports scientific collaborative research.

VI. CONCLUSION

The core argument of this research is that finance should be studied like any hard science, strictly following the procedures of the modern scientific method. We advocate the use of modern computational techniques to support large-scale collaboration to leverage the power of crowds for investigation.

In this paper, we have demonstrated how a trading strategy commonly used in technical analysis could be studied scientifically, for example, with control experiments (Brownian motion), generalization (testing on multiple assets) and statistical analysis. By specifying the experiments under the proposed FRACTI framework, researchers can communicate without ambiguity what experiments they have conducted. As it is the case in hard science, other researchers can repeat the experiments in order to verify the results. This way, FRACTI can support crowd science, which is an efficient way to accelerate research in finance.

In this paper we presented a concrete case in which we use FRACTI to outline a hypothesis, produce, record and collect evidences and get to an objective conclusion about a financial phenomenon we wish to verify.

In this exercise we intentionally selected a model that is simple enough for a wide community of finance users to understand and test. An additional incentive is the exposure of this and similar strategies that are of constant debate, in the media and academia, of opposing views of technical, fundamental and quantitative approaches to investing.

Every single one of the evidences generated in this exercise is called a *contribution*, and as such are shareable artefacts available to a wide community of interested parties. Contributions also inherently carry a number of properties, such as ownership, provenance and access restrictions. These properties are produced and maintained transparently at the same time the contribution is generated.

---

[12] Data snooping, also known as data fishing, data dredging, equation fitting or p-hacking, is the intentional or unintentional use of data inference techniques the researcher decides to perform *after* looking at the data, usually the testing data [67] [68]

[13] Survivorship bias is the unintentional error of concentrating on data items that have "survived" some process and overlooking those that have perished [69]

The investigation exercise core of this paper, described thoroughly in Section V, is evidence of how FRACTI enables "crowd science" [3] through features of a scientific support system described in Section II: transparent collaboration, repeatability of results, accessibility and openness. We explained how these features are required to support "crowd-powered scientific investigation" [1] in large scale in Sections I and II.

A reasonably complex trading strategy defined in Section III can be described by FRACTI models in clear accessible text, for example Figure 6 and Figure 8. By using other researchers' contributions, a researcher does not have to be a computer specialist to understand, communicate and improve them (which is the benefit of crowd science).

A researcher can always use the record of provenance of any FRACTI contributions, as shown in Figure 14, to identify the original data and transformation steps to re-create contributions. The record of provenance allows unquestionable repeatability of results by any researcher who might want to leverage a previous investigation (which is what scientific research should be).

Using FRACTI contributions, we demonstrated how to easily switch datasets between Brownian simulations and real historical data, visualize and back test results – and more importantly – share contributions between interested users and communities (thus supporting collaborative research).

Code examples provided here are not the core subject under research; they are provided for illustration purposes only. Despite that, these features align themselves with the vision of the long-term research and could be looked at as an illustration of the roadmap.

REFERENCES

[1] Jorge M Faleiro Jr and Edward P. K. Tsang, "Supporting Crowd-Powered Science in Economics: FRACTI, A Conceptual Framework for Large-Scale Collaboration and Transparent Investigation in Financial Markets," in *14th Simulation and Analytics Seminar*, Helsinki, 2016.

[2] Raji Al Munir, *A Very Short Introduction to The Modern Scientific Method and The Nature of Modern Science*, Kindle ed., Amazon Books, Ed.: rm@munir.info, 2010.

[3] Chiara Franzoni and Henry Sauermann, "Crowd science: The organization of scientific research in open collaborative projects," *Research Policy*, vol. 43, pp. 1-20, 2014. [Online]. http://ssrn.com/abstract=2167538

[4] Jorge M Faleiro Jr. (2015, Feb.) Representation in Computational Finance. What Really Matters? [Online]. http://bit.ly/1Y68PDw

[5] Jorge M Faleiro Jr. (2008, Aug.) QuantLET: an open source, event-driven framework for real-time analytics. [Online]. http://quantlet.net

[6] Fernando Pérez and Brian E. Granger, "IPython: A System for Interactive Scientific Computing," *Computing in Science and Engineering*, vol. 9, no. 3, May/June 2007, URL: http://ipython.org.

[7] Eric Jones, Travis Oliphant, and Pearu Peterson. (2001) SciPy: Open Source Scientific Tools for Python. [Online]. http://www.scipy.org/

[8] Wes McKinney, "Pandas: Data Structures for Statistical Computing in Python," in *9th Python in Science Conference*, 2010, pp. 51-56.

[9] John Hunter, "Matplotlib: A 2D Graphics Environment," *Computing in Science & Engineering*, vol. 9, pp. 90-95, 2007.

[10] Jorge M. Faleiro Jr. (2015, Jul) QuantLET Example: Backtesting Momentum Strategies using Streams and Monte Carlo Simulations. [Online]. http://goo.gl/EWGqyO

[11] Jorge Martins Faleiro Jr. (2013a) A Scientific Workflow System Applied to Finance.

[12] Jeremy Goecks, Anton Nekrutenko, and James Taylor, "Galaxy: a comprehensive approach for supporting accessible, reproducible, and transparent computational research in the life sciences," *Genome Biology*, Nov. 2010, http://genomebiology.com/2010/11/8/R86.

[13] Thomas Herndon, Michael Ash, and Robert Pollin, "Does High Public Debt Consistently Stifle Economic Growth? A Critique of Reinhart and Rogof," Political Economy Research Institute, University of Massachusetts Amherst, Amherst, Working Paper JEL codes: E60, E62, E65, 2013.

[14] Arie van Deursen and Paul Klint, "Domain-Specific Language Design Requires Feature Descriptions," *Journal of Computing and Information Technology*, vol. 1, pp. 1-17, Oct. 2002.

[15] G., Hohpe and B. Woolf, *Enterprise Integration Patterns*. Boston: Addison-Wesley, 2012.

[16] Jorge Martins Faleiro Jr. (2007, July) Technofinancial Singularity - RMS Architectures. [Online]. http://goo.gl/w7ouS8

[17] Diomidis Spinellis, "Notable Design Patterns for Domain-Specific Languages," *Journal of Systems and Software*, pp. 91–99, February 2001, http://www.spinellis.gr/pubs/jrnl/2000-JSS-DSLPatterns/html/dslpat.html.

[18] Arie van Deursen, Paul Klint, and Joost Visser, "Domain-Specific Languages: An Annotated Bibliography," *ACM SIGPLAN Notices*, vol. 35, no. 6, pp. 26-36, June 2000.

[19] Julien Palard. (2012, Aug.) Pipes: A Infix Notation Library. [Online]. https://github.com/JulienPalard/Pipe

[20] M Schoeffel, "The Three Main Trading Strategies and Their Variations," in *Algorithmic Trading Strategies*, Fudancy Technology, Ed. Nyon, Switzerland: Fudancy Research Group, 2011.

[21] George A Chestnutt, *Stock Market Analysis: Facts and Principles*.: American Investors Service, 1955.

[22] Alfred Cowles, "Can Stock Market Forecasters Forecast?," *Econometrica*, vol. 1, no. 3, pp. 309-324, Jul 1933.

[23] William P Hamilton and Charles H Dow, *The Stock Market Barometer*. New York: Harper & Brothers, 1922.

[24] William P Hamilton. (1903-1929) William Peter Hamilton's Editorials in the Wall Street Jornal. [Online]. http://goo.gl/MoklCh

[25] Scott Patterson, "Technically, a Challenge for Blue Chips," *The Wall Street Journal*, Nov 2007.

[26] Stephen J Brown, William N Goetzmann, and Alok Kumar. (1998, Jan) SSRN. [Online]. http://ssrn.com/abstract=5869

[27] Ben R Marshall, Rochester H Cahan, and Jared M Cahan. (2010, Aug) Techinical Analysis Around the World. [Online]. http://ssrn.com/abstract=1181367

[28] Eugene F Fama, "Random Walks in Stock Market Prices," *Financial Analysis Journal*, pp. 55-59, Sep-Oct 1965.

[29] Robert Brown, *A Brief Account of Microscopical Observations Made in the Months of June, July and August 1827, on the Particles Contained in the Pollen of Plants; and on the General Existence of Active Molecules in Organic and Inorganic Bodies*, Manuscript ed. London, UK: Unpublished, 1827.

[30] V Mandrekar, "Mathematical Work of Norbert Wiener," *Notices of American Mathematical Society*, vol. 42, no. 6, pp. 664-669, Juliy 1995.

[31] Steven W Smith, *The Scientist and Engineer's Guide to Digital Signal Processing*, 1st ed. San Diego, California, USA: California Technical Publishing, 1997.

[32] NIST/SEMATECH. (2012, Apr) e-Handbook of Statistical Methods. [Online]. http://www.itl.nist.gov/div898/handbook//

[33] Eric W Weistein. MathWorld--A Wolfram Web Resource. [Online]. http://mathworld.wolfram.com/TriangularNumber.html

[34] Edward Tsang, "New ways to understand financial markets," Centre for Computational Finance and Economic Agents (CCFEA), University of Essex, Colchester, UK, Working Paper WP046-10 2010.

[35] Kay-Yut Chen, "An Economics Wind Tunnel: The Science of Business Engineering," *Experimental and Behavioral Economics - Advances in Applied Microeconomics*, vol. 13, 2005.

[36] McGraw Hill Financial. S&P 500. [Online]. http://us.spindices.com/indices/equity/sp-500

[37] Michael Friendly and Daniel Denis, "The Early Origins and Development of the Scatterplot," *Journal of the History of Behavioral Sciences*, vol. 41, pp. 103-130, Spring 2005.

[38] McGraw Hill Financial. S&P 500 Fact Sheet. [Online]. http://goo.gl/Ib5AXX

[39] Quandl. (2015, July) Quandl Data Platform. [Online]. http://www.quandl.com

[40] E S Browning, "Reading the Market's Tea Leaves," *Wall Street Journal*, Jan 2006.

[41] Paul A Samuelson, "Proof that Properly Anticipated Prices Fluctuate Randomly," *Industrial Management Review*, vol. 6, no. 2, pp. 41-49, Spring 1965. [Online]. http://www2.math.uu.se/cim/seminars/Samuelson-Proof.pdf

[42] Robert Huebscher. (2009, Jul) Burton Malkiel Talks the Random Walk. [Online]. http://goo.gl/1CPgpr

[43] Christopher Carolan, "Autumn Panics: A Calendar Phenomenon," *Dow Award*, 1998, https://www.mta.org/eweb/docs/1998DowAward.pdf.

[44] Cheol-Ho Park and Scott H Irwin, "The Profitability of Technical Analysis: A Review," Department of Agricultural and Consumer Economics, University of Illinois at Urbana-Champaign, Urbana-Champaign, Research Report 2004. [Online]. http://purl.umn.edu/37487

[45] Cheol-Ho Park and Scott H Irwin, "What Do We Know About the Profitability of Technical Analysis?," *Journal of Economic Surveys*, vol. 21, no. 4, pp. 786-826, Jul 2007.

[46] Nauzer J Balsara, Gary Chen, and Lin Zheng, "The Chinese Stock Market: An Examination of the Random Walk Model and Technical Trading Rules," *The Quarterly Journal of Business and Economics*, Spring 2007.

[47] Arvid I Hoffmann and Hersh Shefrin , "Technical Analysis and Individual Investors," *Journal of Economic Behavior and Organization*, vol. 107, no. November, pp. 487-511, Feb 2014.

[48] Fischer Black, "Noise," *The Journal of Finance*, vol. 41, no. 3, pp. 529-543, July 1986.

[49] Stanley Young and Alan Karr, "Deming, Data and Observational Studies," *Significance*, pp. 116-120, Sep 2011.

[50] Michael Shermer, "How the Survivorship Bias Distorts Reality," *Scientific American*, vol. 322, no. 3, Aug 2014.

[51] Kausik Chaudhuri and Yangru Wu, "Random Walk vs Vreaking Trend in Stock Prices: Evidence from Emerging Markets," *Journal of Banking and Finance*, vol. 27, pp. 575-592, 2003.

[52] T Bisig, A Dupuis, V Impagliazzo, and R Olsen, "The scale of market quakes," *Quantitative Finance*, vol. 12, no. 4, pp. 501-508, 2012.

[53] NPR. (2013, Apr.) Planet Money: How Much Should We Trust Economics? [Online]. http://goo.gl/a0yxiU

[54] Carmem Reinhart and Kenneth Rogoff, "Growth in a Time of Debt," *American Economic Review*, vol. 100, no. 2, May 2010.

[55] Richard Olsen and Clive Cookson, "How Science Can Prevent the Next Buble," *Financial Times*, Feb. 2009.

[56] Jonah Lehrer, "Can We Prevent the Next Bubble?," *Wired*, June 2011.

[57] Edward Tsang. (2014) High-frequency Finance Research Platform, A Wiki-style Global Project. [Online]. http://www.bracil.net/finance/HFF/platform.html

[58] Jorge Martins Faleiro Jr. (2013, May) Implementation of a "Cyclical Long Position Strategy" in QuantLET. Presentation.

[59] T Berners-Lee, R Fielding, and L Masinter, "Uniform Resource Identifier (URI): Generic Syntax," Network Working Group, The Internet Engineering Task Force, RFC STD: 66, 2005.

[60] David Harel and Amir Pnueli, "Statemate: a working environment for the development of complex reactive systems," in *ICSE '88 Proceedings of the 10th international conference on Software engineering*, Los Alamitos, 1988.

[61] Engineer Bainomugisha, Andoni Lombide Carreton, Tom Van Cutsem, Stijn Mostinckx, and Wolfgang De Meuter, "A Survey on Reactive Programming," *ACM Computing Surveys*, vol. 45, no. 4, Aug. 2013.

[62] J. Dean and S. Ghemawat, "MapReduce: Simplified Data Processing on Large Clusters," in *OSDI '04*, San Francisco, 2004.

[63] Stewart Robinson, *Simulation: The Practice of Model Development and Use*. West Sussex: John Wiley & Sons, 2004.

[64] Jorge Martins Faleiro Jr. (2013, Apr.) Reference Model: Architecture, Concepts and Fundaments.

[65] Oxford University, *Oxford Dictionary of English*, 3rd ed., Angus Stevenson, Ed.: Oxford University Press, 2010.

[66] Vic Norton, "Adjusted Closing Prices," Department of Mathematics and Statistics, Bowling Green State University, 2010.

[67] University of Texas. (2011, Nov.) Common Mistakes in Using Statistics. [Online]. https://www.ma.utexas.edu/users/mks/statmistakes/datasnooping.html

[68] George Simth and Shah Ebrahim, "Data Dredging, Bias or Confounding - They Can All Get You Into the BMJ and the Friday Papers," *British Medical Journal*, vol. 325, no. 7378, Dec. 2002.

[69] Michael Schemer, "How the Survivor Bias Distorts Reality," *Scientific American*, Sep. 2014.